\documentclass[12pt]{iopart}

\newcommand{\mdot}{\mbox{$\dot{M}$}}
\newcommand{\msun}{\mbox{$M_{\odot}$}}
\newcommand{\vw}{\mbox{${v_{w}}$}}
\usepackage{graphicx}  
\begin{document}

\title[Turbulence in Bubbles]{Turbulence in Wind-Blown Bubbles around
Massive Stars}

\author{V. V. Dwarkadas}

\address{Dept of Astronomy and Astrophysics, University of Chicago,
5640 S Ellis Ave, AAC~010c, Chicago IL 60637}
\ead{vikram@oddjob.uchicago.edu}
\begin{abstract}

Winds from massive stars ($>$ 8 solar masses) result in the formation
of wind-blown ``bubbles'' around these stars.  In this paper we study,
via two-dimensional numerical hydrodynamic simulations, the onset and
growth of turbulence during the formation and evolution of these
wind-blown ``bubbles''. Our simulations reveal the formation of vortex
rolls during the Main-Sequence stage of the evolution, and
Rayleigh-Taylor instabilities in the subsequent stages due to
accelerating and/or decelerating wind-blown shells. The bubble shows a
very turbulent interior just prior to the death of the star, with a
significant percentage of the internal energy expended in non-radial
motions. This would affect the subsequent evolution of the resultant
supernova shock wave. We discuss the implications of these results,
show how the ratio of kinetic energy in radial versus non-radial
motions varies throughout the evolution, and discuss how these results
would carry over to three dimensions.

\end{abstract}

\maketitle

\section{Introduction}
Mass loss from stars is a ubiquitous process. During their lifetime,
stars lose mass mainly via stellar winds. These winds may be driven by
coronal pressure in stars like the sun, by radiation pressure on dust
grains such as in Asymptotic Giant Branch stars, or by radiation
driving due to line or continuum opacity in massive stars (for further
details see Lamers and Casinelli 1999). Mass loss via winds is mainly
characterized by the mass-loss rate and velocity of the wind, and
these parameters may change continuously throughout the stellar
lifetime (Garcia-Segura et al.~1996; Langer et al.~1994).

The interaction of the wind from the star with the ambient medium
leads to the formation of wind-blown ``bubbles''. The structure and
morphology of these wind-blown bubbles was first elucidated in a
seminal paper by Weaver et al.~(1977), and has been explored by
several authors since. While the model has been hugely successful in
explaining the global structure of wind-blown bubbles, the finer
details can only be explored by multi-dimensional numerical
simulations. These have been carried out by several authors (for
recent results see Freyer et al 2006; Arthur 2007; Dwarkadas 2007b;
van Marle et al 2007). These authors have examined various different
aspects of the problem. The first two included the ionization from the
star in their numerical calculations, but assumed that the wind
parameters were constant in the various phases. Dwarkadas (2007b) did
not include the effects of stellar ionization, but took into account
the continuous variation of the wind parameters by adopting them
directly from the output of a stellar evolution code. Unlike a
previous calculation (Garcia-Segura et al.~1996), Dwarkadas (2007b)
ran the simulations in 2-dimensions from the beginning. In order to
take the evolution in the wind parameters into account accurately, the
simulation was run for almost 2 million timesteps, one reason why it
is computationally intensive to include the time evolution along with
the ionization effects.

The variation in the wind parameters results in considerable
turbulence within the bubble interior, which was not seen in the work
of Arthur (2007) and Freyer et al.~(2006) as they assumed average (and
constant) wind parameters in each stage of evolution. In Dwarkadas
(2007b) we touched upon some aspects of this turbulence, but
concentrated mainly on the hydrodynamical aspects of the wind-bubble
evolution and the subsequent evolution of a SN shock wave within the
wind bubble. In this companion paper we explore in more detail the
formation and growth of turbulence within the wind blown bubble, and
the implications for the subsequent evolution, the emission from the
bubble and observations of wind-blown bubbles, and future 3D
simulations. Our aim is not to explore the quantitative aspects of
turbulence or derive energy spectra. In keeping with the objectives of
the conference where this work was presented, they are to explore the
manifestation of turbulence and turbulent mixing mechanisms in
wind-blown bubbles around massive stars, accentuate the factors that
lead to turbulencein these conditions, and explore its effects on the
hydrodynamics and subsequent evolution.

The rest of this paper proceeds as follows: In \S \ref{sec:windbub} we
describe the basic structure and evolution of a wind blown bubble. In
\S \ref{sec:numsim} we give a brief description of wind-blown bubbles
around massive stars. In \S \ref{sec:turbulence} we examine in detail
the various manifestations of turbulence that we see in our
simulations. Finally \S \ref{sec:summary} discusses the implications
of our results and prospects for future work.

\section{The Structure of a Wind-Blown Bubble}
\label{sec:windbub}
The general structure of a wind-blown nebula was first elucidated by
Weaver et al.~(1977).  In the simplest, two-wind approximation, a fast
wind from a star collides with slower material emitted during a
previous epoch, driving a shock into the ambient medium. The pressure
of the post-shock material causes the freely flowing fast wind to
decelerate, driving a second shock that propagates into the wind. A
complex double-shocked structure is formed, separated by a contact
discontinuity. Figure 1 shows the density and pressure profiles from a
simulation of a wind-blown bubble. Going outwards in radius from the
central star we find the following regions delineated: freely flowing
fast wind, inner or wind-termination shock (R$_t$), shocked fast wind,
contact discontinuity (R$_{cd}$), shocked ambient medium, outer shock
(R$_o$) and unshocked ambient medium.

\begin{figure}
\includegraphics{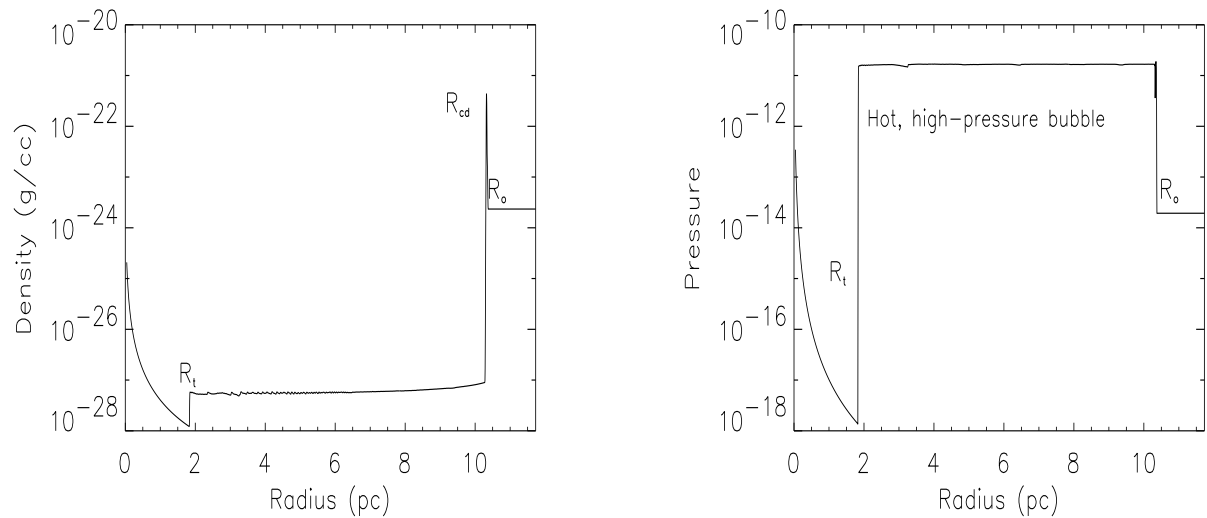}
\caption{\small a) Density and b) Pressure profiles from a numerical
simulation of a wind-blown bubble around a massive star. }
\end{figure}

The shocked ambient medium is cooled and compressed into a thin, dense
shell, which we will denote throughout as the circumstellar
shell. Much of the volume of the bubble is occupied by a high
pressure, low density region. The temperature within this region can
be as high as 10$^7$ to 10$^8$K. One would therefore expect
significant X-ray emission from the bubble. However diffuse X-ray
emission from nebulae surrounding massive Wolf-Rayet stars has been
seen only in very few objects (Chu \etal 2006).

The assumption of constant wind properties is not a very good one. As
we have said before the wind properties, which can be thought of as a
function of the stellar properties, evolve with the star. Evolving
winds may change the basic structure considerably. In particular, as
we show in \S 4, they may introduce multi-dimensional effects such as
the presence of hydrodynamic instabilities and the onset of turbulence
(see also Dwarkadas \& Balick [1998] for the planetary nebulae
case). However even when the wind is evolving, as long as its
mechanical luminosity (0.5$\mdot {\vw}^2$) remains constant, where
$\mdot$ is the mass-loss rate and $\vw$ is the wind velocity, the
above profiles will remain more or less valid. A major change in the
pressure equilibrium is needed to alter the above profiles, which
generally happens when stars go from one phase to another.

\section{Brief Overview of  Wind-Blown Bubbles around Massive Stars}
\label{sec:numsim}

Arthur (2007) has provided an elegant description of the formation of
the wind-blown medium around a 40 $\msun$ star, including the effects
of the ionizing photons from the star. Dwarkadas (2005, 2007a,b,c) has
summarized the properties of the bubble during various stages of the
evolution and for stars of various initial masses. A wonderfully
comprehensive review on the evolution of massive stars is given in
Woosley et al.~(2002). Herein we will only very briefly summarize some
of these findings.

A massive star begins its life in the main-sequence (MS) stage, as an
early type O or B star. These stars lose mass mainly through radiative
driving, due to opacity in various lines. The wind has a mass-loss
rate of about 10$^{-8}$ to 10$^{-6} \msun$/yr, and a wind velocity of
2000-4000 km/s. This is the longest, hydrogen burning phase, which can
last for a few million years. The star will be expected to form a
massive bubble upto several tens of parsecs in radius. As explained in
\S \ref{sec:windbub}, a double shocked structure is formed, with the
inner shock generally only a few parsecs away from the center.

The subsequent evolution generally occurs within this bubble. Stars
with solar metallicity will usually become red supergiants (RSG), once
the hydrogen burning is completed. The star will swell up in size to
greater than about 10$^{13}$ cm, and begin to burn helium in the
core. The large size implies a much smaller escape velocity off the
surface of the star, and therefore a smaller wind velocity. RSG winds
are assumed to be driven by radiation pressure on dust grains. The
measured velocities are very slow, on the order of $v_w \sim 10-50$ km
s$^{-1}$, and the mass loss rates are quite high, approaching $\mdot
\sim 10^{-4} \msun$/yr. Due to the low velocity, the RSG wind does not
travel very far. The wind density (dependent on $\mdot/v_w$) is very
high because of the low velocity, and thus it forms a high density
region with a new pressure equilibrium. For constant wind parameters,
the density decreases as r$^{-2}$.

Stars less than about 30-35 $\msun$ will end their lives as
RSGs. Stars greater than this mass may go on to form Wolf-Rayet (W-R)
stars, which have very fast radiatively driven winds. Very massive
stars ($> 50 \msun$ ) may also experience an unstable Luminous Blue
Variable phase, wherein they ejecta a large amount of material over a
short period of a few years, giving rise to an LBV bubble within the
MS bubble.

Dwarkadas (2007b) completed a thorough study of the evolution of a 35
$\msun$ star that starts off as a MS star, goes through the RSG stage
and then forms a W-R bubble.  We refer the reader to that paper (and
Freyer et al.~2006) for a detailed study of the hydrodynamics of the
evolution. Herein, we will describe pertinent details of the
hydrodynamics for the sake of completeness, while concentrating on
aspects of turbulence that were not included in the earlier paper.

\section{Turbulence in Wind-Blown Bubbles}
\label{sec:turbulence}

In this paper we concentrate on two-dimensional simulations of the
formation of wind-blown bubbles. These simulations were carried out
with the VH-1 code, a 1, 2 and 3D numerical hydrodynamics code (see
Blondin \& Lundqvist [1993] for a more detailed description) that
solves the equations of mass, momentum and energy conservation on a
uniform grid. The calculations are carried out on a spherical ($r -
\theta$) grid in a Lagrangian frame, with the results at each timestep
being remapped to the original Eulerian grid. Radiative cooling is
included via a cooling function. A feature of this code is that it
includes an expanding grid, which tracks the outer shock and expands
outwards with it. This feature is exceptionally useful in situations
such as the current one where the object being simulated expands
almost 5 orders of magnitude

In our simulations of wind-blown bubbles we see various manifestations
of turbulence. The initial {\bf main-sequence stage} shows significant
evidence of vorticity being deposited into the interior (shocked wind)
region of the flow. This vorticity arises at the inner,
wind-termination shock region. It is due to the slight change in the
position of the termination shock (with respect to the outer shock and
to its previous position) at every timestep, or every few
timesteps. The change in the position leads to a change in the
pressure and density just behind the wind termination shock. The
post-shock flow, being subsonic, does not have enough time to
equilibrate before the shock position changes again. This leads to
continual changes in the pressure, which consequently leads to changes
in the velocity and the development of a velocity gradient behind the
inner shock. $\nabla \times v$ becomes non-zero, and vorticity is
deposited within the interior flow.

Although it is clear why the position of the wind termination shock
changes (due to changes in the wind properties), it is not as clear
why the shape of the shock does. It is possible that this starts
mainly from near the azimuthal axis, and is due to the so-called axis
effect in a 2-D simulation. This is due to the fact that flow that is
at an angle to the azimuthal axis is not allowed to cross the axis but
is forced to slide along the axis in a 2-D calculation. The wind shock
at the axis position is therefore extended or depressed, depending on
whether the flow is moving away from the pole or towards it. The
change in shape leads to changes in the radial density and pressure.

It is debatable whether such a situation would occur in the 3D case,
when their is no artificial boundary and the flow would be able to
cross the azimuthal axis. The important question however is whether a
3D situation would exist that would result in a wind termination shock
that is not completely spherical. The answer to this is a qualified
yes. The shape of the shock front responds to inhomogeneities in the
flow, the presence of clumps and other density perturbations, or
hydrodynamical instabilities. A case in point is the shock fronts seen
in simulations of planetary nebulae (Dwarkadas \& Balick [1998], see
Fig 2) where the shape of the shock front responds directly to
finger-like clumps in the flow, and results in box-shaped inner
shocks. As long as there is some entity capable of introducing
instability within the flow, which is likely as long as the wind
parameters are constantly changing, the result would be a distortion
of the shock front and production of vorticity.

What is the result of this vorticity deposition? The vorticity is
carried out with the shocked wind, and vortex rolls slowly fill the
shocked wind. A snapshot of the velocity flow in the region is shown
in Fig 2. The formation and clustering of vortices in the hot bubble
is apparent. The vortices last for a long time, some of them lasting
over the remaining main sequence lifetime since their birth. They also
grow in size over the simulation. This is not visually apparent
because the grid is expanding in size, but the fact that they retain
their relative size as the grid grows over several orders of magnitude
indicates the vortex growth. Thus, as is commonly seen in 2D
simulations as opposed to the 3D case, the energy tends to cascade to
longer wavelengths. There is a definite clustering of vortices, and
vortex rolls near the axis tend to merge together. The overall effect
is that some amount of energy is expended in non-radial turbulent
motions.

\begin{figure}
\includegraphics[scale=0.9]{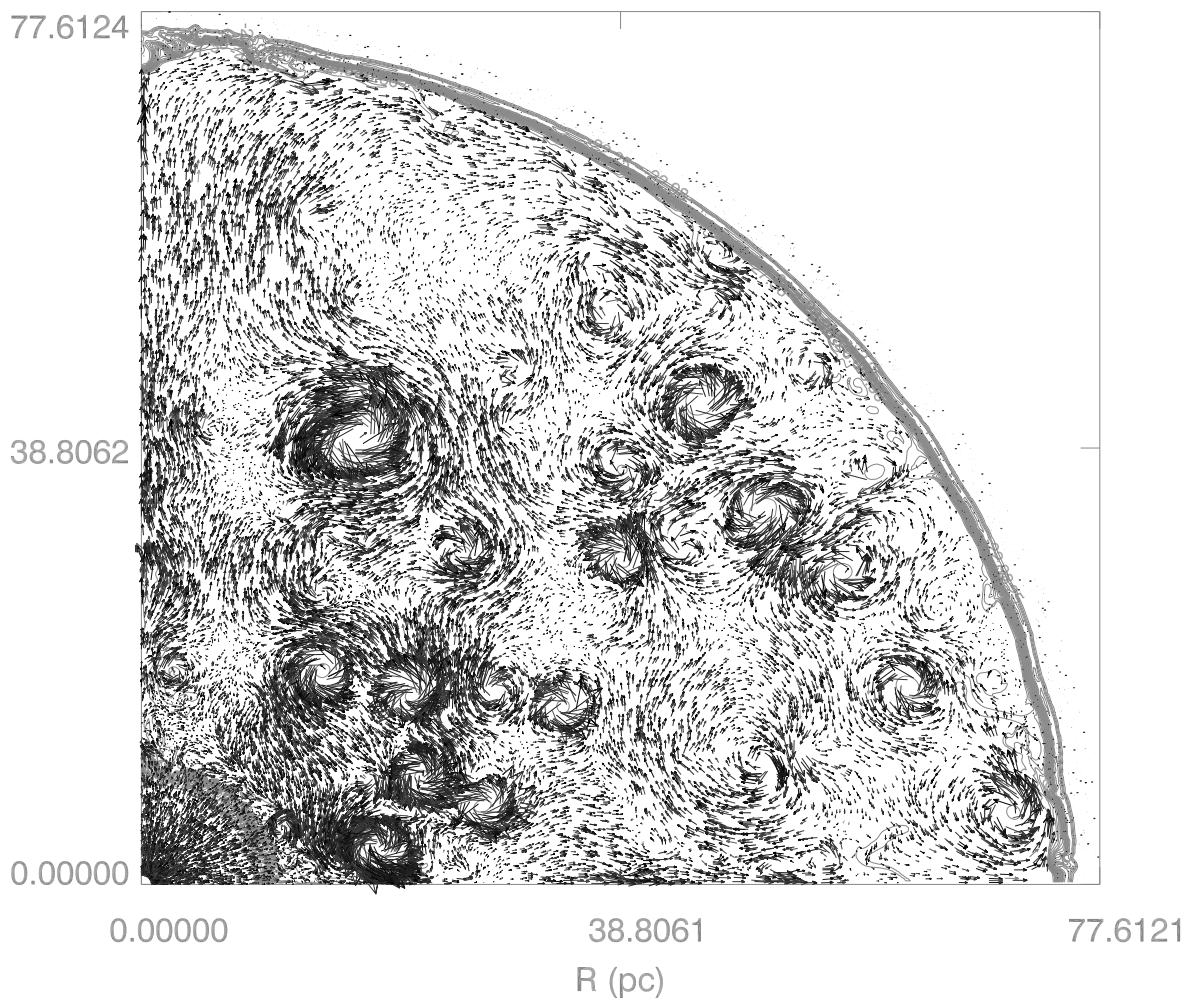}
\caption{\small A snapshot showing the velocity contours during the
main-sequence stage of the evolution of the wind-blown bubble. The
formation and clustering of vortices in the shocked wind is clearly
visible}
\end{figure}

In order to calibrate how much of the energy is expended in such
motions, we calculated for each zone the kinetic energy per unit
volume of non-radial motions, and divided by the total energy per unit
volume. We find that a few percent of the energy, typically 3-5 \%, is
expended in non-radial motions, although at times it can
instantaneously be up to 7-8 \%. We also find that this fraction is
more or less constant over varying resolution, showing that numerical
resolution is not playing a large effect.

In the main sequence stage we see turbulence within the shocked
wind. The onset of the {\bf RSG stage} in our calculation leads to the
development of a new pressure equilibrium and a thin RSG shell which
is decelerated by the strong thermal pressure in the interior as it
expands outwards. The decelerating shell is found to be unstable to
the Rayleigh-Taylor (R-T) instability, and Rayleigh-Taylor fingers are
seen developing that expand outwards from the high density gas into
the low density external medium. This situation is shown in Fig 3,
where density contours (grey shading) are shown overlapping the
velocity vectors. Note that this phase lasts for a much shorter time,
about 250,000 years, compared to the Main Sequence phase which last
for about 4.5 million years. The presence of the R-T instability leads
to some turbulence around the radius where the RSG material is piled
up in a thin shell, but it is localized, and does not have sufficient
time to grow because the duration of this phase is small.

\begin{figure}
\includegraphics[scale=0.82]{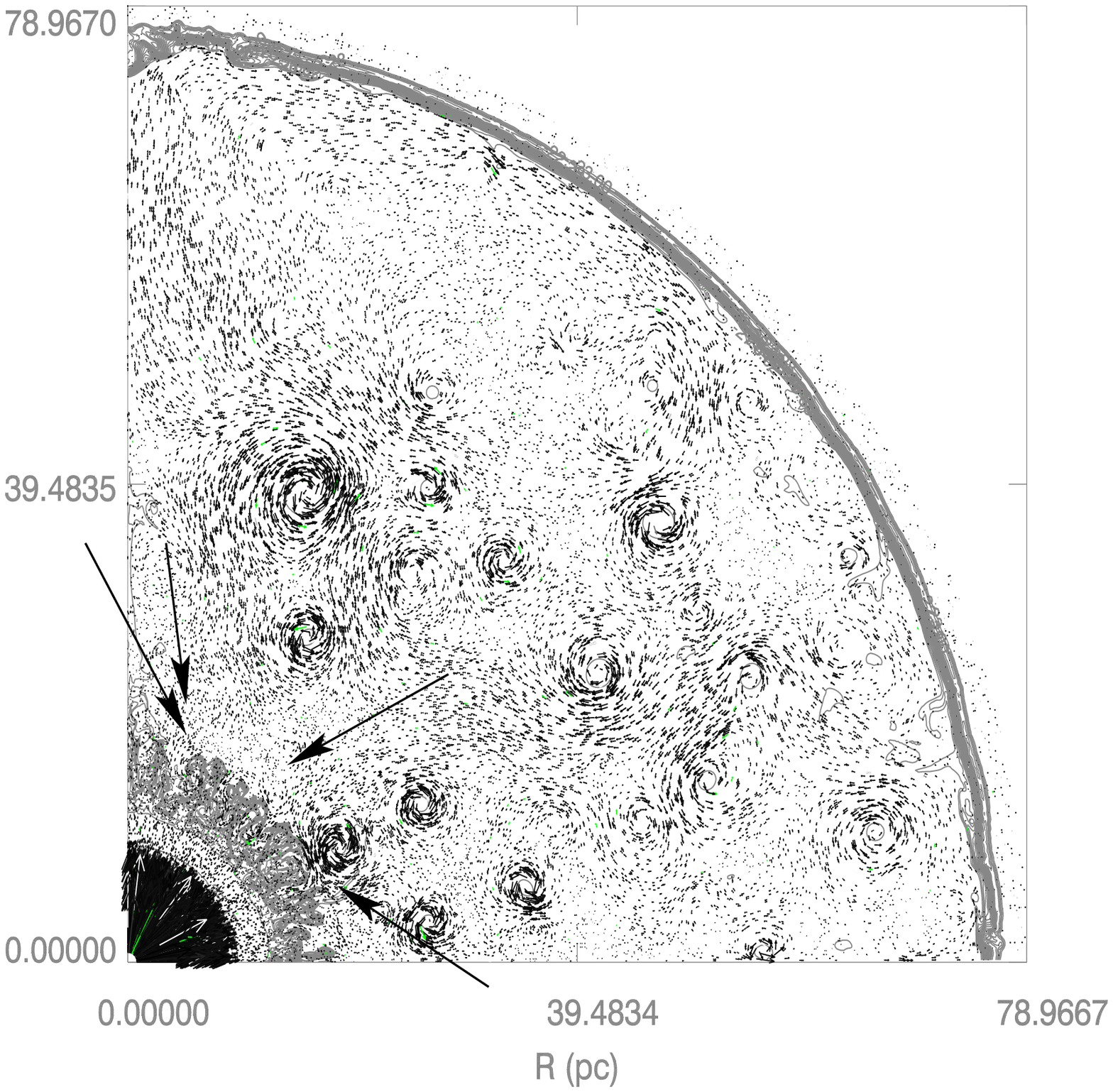}
\caption{\small A snapshot showing the velocity contours during the
red-supergiant stage of the evolution of the wind-blown bubble, with
density contours superimposed. The arrows point to the Rayleigh-Taylor
unstable thin shell of material, about 11pc from the center (1 pc = 3
$\times 10^{18}$ cm.)}
\end{figure}

The star then becomes a hot, blue, compact {\bf Wolf-Rayet} (W-R)
star. The wind velocity increases by almost two orders of magnitude,
the mass-loss rate drops by a few, and the wind momentum therefore
increases considerably. The W-R wind accelerates down the RSG density
incline, the high pressure accelerating the dense W-R shell.  The W-R
shell is also therefore unstable to R-T instabilities with a very
different Atwood number from the RSG case. However the dense region
being accelerated is now exterior to the low density, high pressure
region that is accelerating it. The R-T fingers therefore expand
inwards from the dense shell into the interior of the W-R bubble. Our
simulations clearly show the growth of inward-pointing protrusions in
the W-R shell. Again however due to the smaller time period of the W-R
phase (about 200,000 years) and the resolution of our simulations, the
growth of the R-T fingers is limited.

The high momentum W-R wind succeeds in breaking up the RSG shell,
which is already fragmented as described earlier, and distributing the
material within the interior. The supersonic W-R material carries the
RSG material along with it. However instead of a shell of material
expanding out, it is a disjointed flow. Note that the amount of mass
lost in the RSG stage is about 19.5 solar masses, while in the W-R
stage it is only a couple of solar masses. So the high momentum W-R
wind basically carries mostly RSG wind material. This material impacts
the main-sequence shell and bounces back, but not in an ordered
fashion, resulting in a highly turbulent medium at the end of the
simulation. A snapshot of this medium just a few thousand years before
the death of the star is shown in Fig 4. The bounce of the W-R wind
from the main-sequence shell, and the subsequent formation of an inner
shock where the pressure of the wind equals the post-shock pressure
behind the reflected material is seen. What is also apparent is the
stirring up of the medium beyond the inner shock, due to the passage
of the expanding W-R wind and the reflection from the outer shock.

\begin{figure}
\includegraphics[scale=0.82]{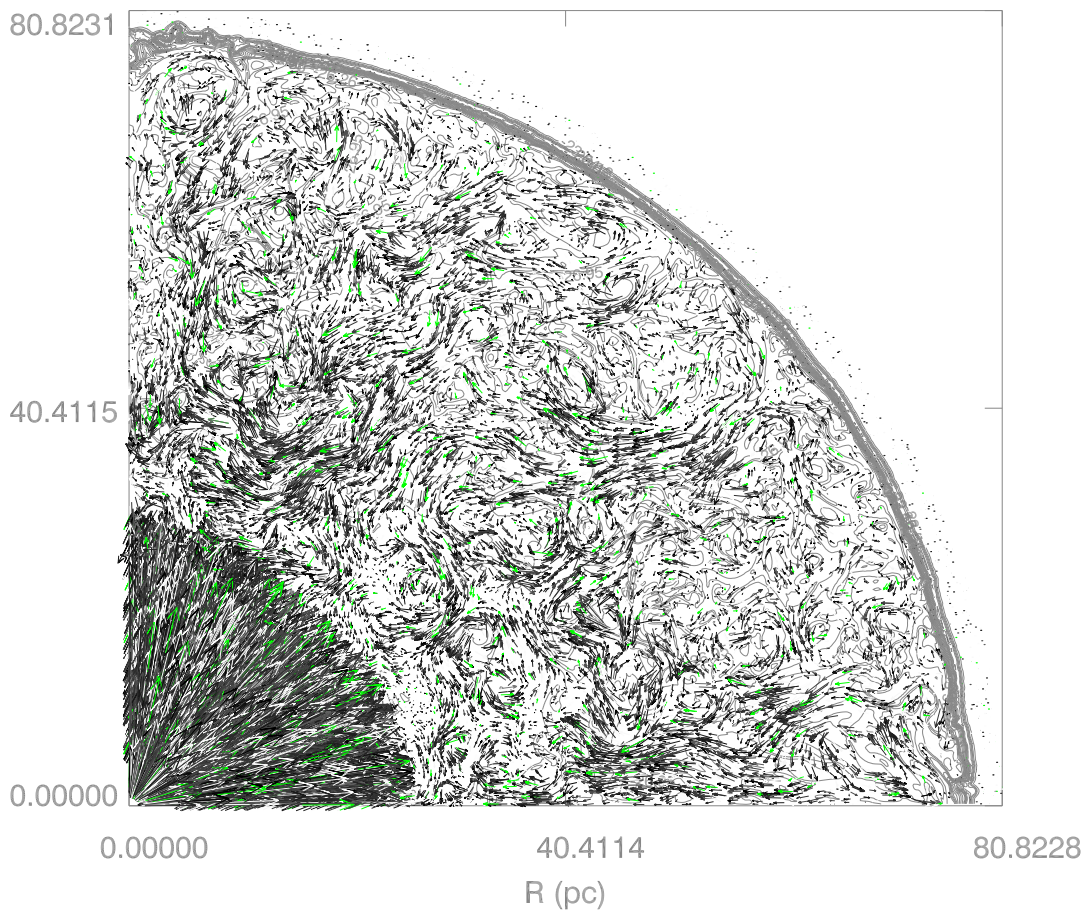}
\caption{\small A snapshot showing the velocity contours during the WR
stage of the evolution of the wind-blown bubble. The WR wind has
broken through the RSG shell, carried the material out to the MS
shell, collided with it and reflected back. This results in a very
turbulent interior within the cavity. }
\end{figure}

The ratio of the energy in non-radial motions to that in radial
motions is constantly changing throughout this evolution. In Fig 5 we
show how the ratio of the energy in non-radial motions to that in
radial motions varies through the simulation. We measure this ratio in
the low density interior between the reverse shock and the contact
discontinuity. For most of the MS phase the value is only
2-3\%. Towards the end of the MS phase, the last 0.5 million years or
so, this ratio increases to 4-5\%.  It decreases in the RSG phase as
expected, because the RSG region does not stretch too far out. But
when the W-R wind impacts the RSG shell and breaks it apart, mixing
the material into the surrounding region, this ratio increases
considerably to almost 20\%. At that point almost 1/6th of the kinetic
energy in the interior is going into non-radial motions.

\begin{figure}
\includegraphics[scale=0.65, angle=90]{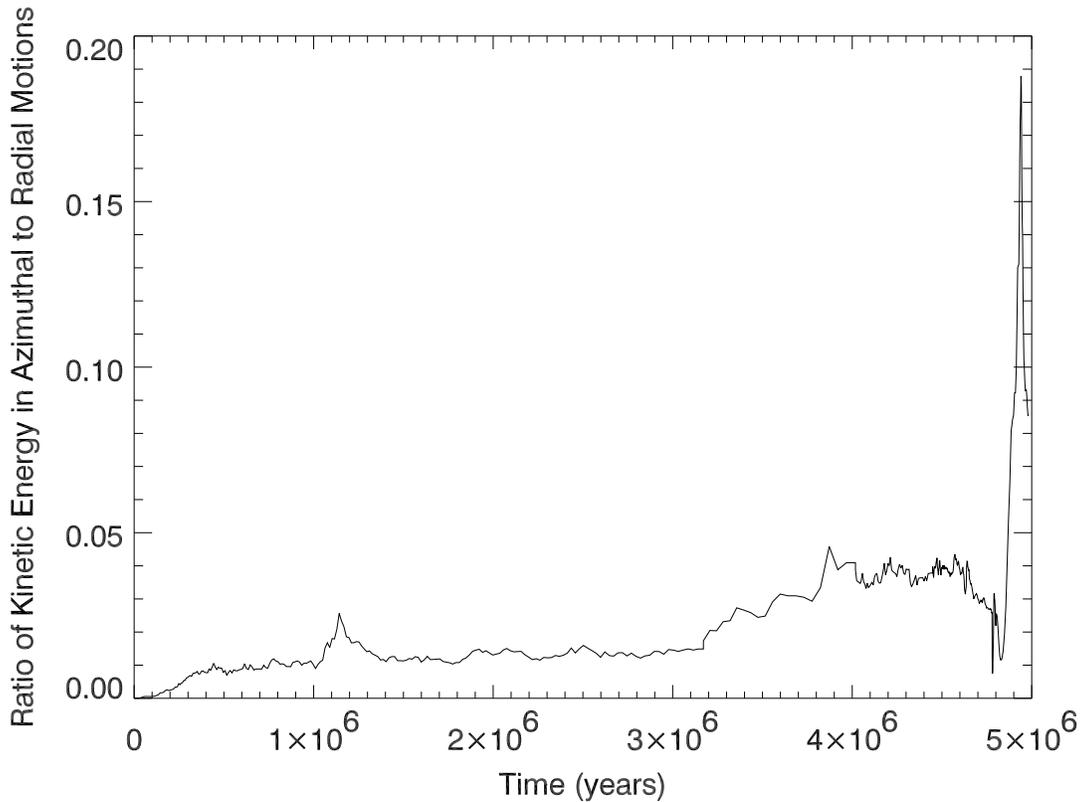}
\caption{\small The ratio of energy in non-radial motions to that in
radial motions in the interior low density cavity, throughout the
stellar lifetime. The ratio starts out low (0.01-0.02) as expected,
increases over the MS stage to 0.04-0.05, drops in the RSG stage and
then increases significantly, to almost 0.2, in the WR phase.}
\end{figure}

\section{Discussion, Summary and Future Prospects}
\label{sec:summary}

Turbulence implies various things to people. To many it signifies a
perturbation, generally non-linear, that disrupts a smooth background
flow. Elmegreen and Scalo (2004) in a comprehensive review article,
define it as ``non-linear fluid motion resulting in the excitation of
an extreme range of correlated spatial and temporal scales.'' Unstable
regions caused by the growth of hydrodynamical instabilities can
disrupt the smooth background flow, sometimes referred to as turbulent
behavior. Turbulence can also mean the complete disruption of the
flow, for example a radial flow showing considerable non-radial
motions, the presence of vortices and strong inhomogeneities in the
flow properties. Turbulence generally implies a high Reynolds number,
but the Reynolds number in simulations is many orders of magnitude
below those seen in nature.

In this paper, we have investigated the turbulence generated during
the evolution of a wind bubble around a massive star. We find that
turbulence within the interior is reflected in the formation of
vortices during the star's main-sequence phase, and in the growth and
evolution of Rayleigh-Taylor instabilities during the RSG and W-R
phases. The turbulence reaches a maximum close to the end of the
star's lifetime, when the WR shock completely breaks up the fragmented
RSG shell, carries the material out till it impacts the MS shell and
reflects back from it. The instabilities in the RSG and WR phases, the
fragmentation of the RSG shell by the WR wind and the resultant uneven
bounce from the MS shell lead to the turbulent interior of the bubble
at the end of the star's life, when it will collapse in a massive SN
explosion. The resulting SN shock wave will expand into this turbulent
medium.

The amount of energy expended in non-radial, turbulent motions is
important for astrophysical purposes. The internal energy in the hot
shocked wind region leads to a very high temperature in that
region. The standard Weaver et al.~(1977) model predicts a temperature
of about 10$^7$ to 10$^8$ K, which would result in strong diffuse
X-ray emission. However, very few windblown nebulae are seen in
diffuse X-ray emission (Chu et al.~2003, 2006; Wrigge et al.~2005).
One suggestion is that the amount of emitting material is very low, as
the densities in the interior are very low. While that is a
possibility, it still is insufficient to explain the very low
temperatures that are seen, on the order of 1.5 $\times 10^6$ K. This
suggest that some entity is reducing the temperature within the
bubbles. If the post-shock kinetic energy were used up in turbulent
motions and not available as internal energy, then this would reduce
the temperature. However this would require a much larger percentage
of the energy to be expended in turbulent motions than we
observe. Thus turbulence alone can certainly not resolve this
problem. However it is possible that turbulence, along with thermal
conduction, mixing and hydrodynamical instabilities and other factors
may all contribute to lowering the internal temperature.

Besides its importance in the dynamics of the wind-bubble itself, the
turbulence has a significant impact on the evolution of the subsequent
SN shock wave, as comprehensively shown in Dwarkadas (2007b). The
interaction of the SN shock wave with the turbulent medium, and
interaction with high-density clumps, results in wrinkling of the
shock wave, as some parts get compressed while others expand
faster. The corrugated shock then impacts the main-sequence shell over
a period of time rather than all at once. Each impact will cause an
increase in the optical and Xray emission from the source. Therefore
this interaction will cause the dense shell to brighten at different
points at different times, giving a ``Christmas-lights''
appearance. Such a phenomenon, of bright spots appearing around a ring
(rather than a shell) due to a SN shock front impacting it has been
seen in SN 1987A. Although these simulations are by no means intended
to explain SN 1987A, our simulations do provide a reasonable guide to
at least deciphering some of what may be going on in SN 1987A. Further
simulations are urgently needed to confirm the existence of the
turbulent interior in wind blown bubbles around stars of various
initial mass undergoing various evolutionary scenarios.

The simulations presented in this paper were carried out on a
two-dimensional grid. It is well known that 2D turbulence has very
different properties from 3D turbulence. 2D turbulence results in an
inverse cascade in energy, as opposed to 3D turbulence, where energy
cascades to lowest scales (Smith \& Yakhot 1997; Danilov \& Gurarie
2000; Delbende et al.~2004; Scalo \& Elmegreen 2004; Bruneau et
al.~2007). It has been shown that 2D turbulence results in a
clustering of vortices which may be long-lived, and the behavior is
not reproduced in 3D. Our simulations do show these aspects, although
we do not see merging of vortices except near the inner shock
region. All these results, though well documented, are based on
experiments conducted for 2D incompressible flows. Our simulations
deal with 2D axially symmetric compressible turbulence. It is not easy
to decide to what extent the results will carry over, although
undoubtedly they will be applicable at some level. Therefore the
formation of the vortices in the main sequence phase, and the
turbulent nature of the shocked wind in that phase, may be
questionable. In the later phases, the formation and growth of the R-T
fingers will still occur. Their length and structure will be different
in the 3D calculations as compared to the 2D ones, but these
differences are quantitative, rather than qualitative.

In the final analysis, it is clear from Fig.~5 that the essential
result, of the turbulence within the bubble at the end of the star's
lifetime, is not dependent on the main-sequence evolution so much as
the post-main sequence phases. The formation of instabilities, and in
particular the collision of the W-R wind with the RSG wind, the
fragmentation of the RSG shell, and the impact with the MS shell to
give the mixing and stirred up interior, is as likely to occur in 3D
as it did in 2D, although the quantitative results may be slightly
different. We therefore assert that the important qualitative results
presented herein will still prove robust in 3D simulations. We are
currently preparing 3D runs using the astrophysical hydrodynamics code
FLASH (Fryxell et al.~2000) with which we will test this assertion,
and will report those results in a future paper.

{\bf Acknowledgements:} VVD is supported by award \# AST-0319261 from
the NSF. I am grateful for support provided by the TMBW07 organizing
committee that enabled me to attend the conference. I would like to
thank the conference organizers, and especially Dr.~Abarzhi, for an
excellent meeting in a beautiful location.

\References

\item[]Arthur S J 2007 {\it RMxAC} {\bf 30} 64

\item []Blondin, J M and Lundqvist P 1993, {\it ApJ}, {\bf 405}, 337

\item []Bruneau C-H, Fischer P and Kellay H 2007, {\it EPL}, {\bf 78},
34002

\item[]Chu Y-H, Gruendl R A and Guerrero M A 2006 {\it The X-ray
Universe 2005} Ed A Wilson vol 1 (Noordwijk: ESA Publications
Division) p~363

\item[] Chu Y-H, Guerrero M A et al.~2003, {\it ApJ}, {\bf 599}, 1189

\item []Danilov S D and Gurarie D 2000, {\it
UsFiN}, {\bf 43}, 863

\item [] Delbende I, Gomez T, Josserand C, Nore C and Rossi M 2004,
{\it CRM}. {\bf 332}, 767

\item [] Dwarkadas V V 2007a {\it ApSS}, {\bf 307}, 153

\item [] Dwarkadas V V 2007b {\it ApJ} {\bf 667} 226

\item [] Dwarkadas V V 2007c {\it RMxAC} {\bf 30} 49

\item [] Dwarkadas V V, 2005, {\it ApJ}, {\bf 630}, 892

\item [] Dwarkadas V V and Balick B 1998, {\it ApJ}, {\bf 497}, 267

\item [] Elmegreen B G and Scalo J 2004, {\it
ARAA}, {\bf 42}, 211

\item [] Freyer T, Hensler G and Yorke H W 2006, {\it
ApJ}, {\bf 638}, 262

\item [] Fryxell, B, Olson K, Ricker P et al 2000,
{\it ApJS}, {\bf 131}, 273

\item [] Garcia-Segura G, MacLow M.-M, \&
Langer N 1996, {\it A\&A} {\bf 305}, 229

\item [] Lamers, H J G L M, and Casinelli, J P
1999, {\it Introduction to Stellar Winds}, (Cambridge: Cambridge
University Press)

\item [] Langer N, Hamman W-R, et al 1994, {\it A\&A},
{\bf 372}, 819

\item [] Scalo J and Elmegreen B G 2004, {\it
ARAA}, {\bf 42}, 275

\item [] Smith L M and Yakhot V 1997, {\it PR},
{\bf 55}, 5458

\item [] van Marle A J, Langer N and García-Segura,
G 2007, {\it A\&A} {\bf 469} 941

\item [] Weaver R, McCray R, Castor J, Shapiro P, \& Moore R 1977 {\it
ApJ} {\bf 218} 377

\item [] Woosley S E, Heger, A, and Weaver, T A 2002, {\it RMP}, {\bf
74}, 1015

\item[] Wrigge M, et al.~2005, {\it ApJ}, {\bf 633}, 248

\endrefs

\end{document}